\documentclass[useAMS,usenatbib]{mn2e}
\usepackage{graphicx}

\title[The redshift distribution of SMGs]{The redshift distribution  of submillimetre galaxies at different wavelengths}
\author[J. A. Zavala, I. Aretxaga and D. H. Hughes]{J. A. Zavala\thanks{E-mail: zavala@inaoep.mx},  I. Aretxaga and D. H. Hughes\\
Instituto Nacional de Astrof\'{i}sica, \'{O}ptica y Electr\'{o}nica (INAOE), Luis Enrique Erro 1, Sta. Ma. Tonantzintla, Puebla, Mexico}

\begin{document}

\date{Accepted 2014 June 26. Received 2014 June 25; in original form 2014 April 1}
\volume{443} \pagerange{2384 -- 2390} \pubyear{2014}

\maketitle

\label{firstpage}

\begin{abstract}
Using simulations we demonstrate that some of the published redshift
distributions of submillimetre galaxies (SMGs) at different
wavelengths, which were previously reported to be statistically
different, are consistent with a parent distribution of the same
population of galaxies. The redshift distributions which peak at
$z_{med}=2.9$, 2.6, 2.2, 2.2, and 2.0 for galaxies selected at 2 and
1.1 mm, and 870, 850, and 450 $\mu$m respectively, can be derived from a single 
parent redshift distribution, in contrast with previous studies. The
differences can be explained through wavelength selection, depth of
the surveys, and to a lesser degree, angular resolution. The main 
differences are attributed to the temperature of the spectral energy
distributions, as shorter-wavelength maps select a hotter population
of galaxies. Using the same parent distribution and taking  into account  
lensing bias we can also reproduce the redshift distribution of 
1.4 mm-selected ultra-bright galaxies, which peaks at $z_{med}=3.4$. However,  
the redshift distribution of 450 $\mu$m-selected galaxies in the deepest surveys,
which peaks at $z_{med}=1.4$, cannot be reproduced from the same parent population 
with just these selection effects. In order to explain this distribution we have to
add another population of galaxies, or include different selection biases.
\end{abstract}


\section{Introduction}

The discovery of a large population of bright sources at high redshift
through the opening of the submillimetre (submm) and millimetre (mm)
wavelength windows (e.g. \citealt{1997ApJ...490L...5S};
\citealt{1998Natur.394..241H}) continues to have a profound impact on
our understanding of galaxy evolution in the early Universe. These
submm/mm-selected galaxies (hereafter SMGs) are characterized by large
far-infrared (FIR) luminosities ($\ga10^{12}$ L$_\odot$), tremendous
star formation rates (SFRs, $\ga300$ M$_\odot$ yr$^{-1}$), large gas
reservoirs ($\ga10^{10}$ M$_\odot$), and a number density that is high
compared to local ultra-luminous infrared galaxies (see reviews by
\citealt{2002PhR...369..111B} and \citealt{2014arXiv1402.1456C}).

Due to the steep rise with frequency of the spectral energy
distribution (SED) of this population of galaxies on the
Rayleigh--Jeans tail ($S_\nu\propto\nu^{3-4}$), the FIR peak is
redshifted into the sub-mm/mm observing bands with increasing
distance, resulting in a strong negative $k$-correction that roughly
cancels the effects of cosmological dimming with redshift for
observations at $\lambda\ge500 \mu$m and within $1\la z\la10$
(\citealt{1993MNRAS.264..509B}). This effect represents a unique
opportunity for an unbiased view of star formation over a wide
redshift range back to the earlier epochs of structure
formation. However, identifying and understanding the nature of these
discrete sources has proven to be challenging because of the low
angular resolution of single-dish telescopes ($\sim14-35$ arcsec), the
faintness of counterparts in the rest-frame optical and ultraviolet
bands, and the limited statistics of poor samples
(\citealt{2002PhR...369..111B}, and references therein). Significant
effort, using multi-wavelength observations to identify counterparts,
has been made to calculate the redshift distribution of SMGs.
The use of high resolution radio continuum and {\it 
Spitzer}/Multiband Imaging Photometer for {\it Spitzer} (MIPS) 24
$\mu$m images suffers from a well-known systematic bias against high
redshift (z $\ga$ 3) sources. Indeed, a large fraction of the
counterparts identified using direct interferometric imaging in the
mm/submm wavelengths are shown to be extremely faint in nearly all
other wavelength bands (r $>$ 26, K $>$ 24) with little or no radio or
{\it Spitzer}/MIPS 24 $\mu$m emission (\citealt{2006ApJ...640L...1I};
\citealt{2007ApJ...670L..89W}; \citealt{2007ApJ...671.1531Y};
\citeyear{2009ApJ...704..803Y}), and a fraction of high redshift SMGs may have been
missed or mis--identified with a foreground source in earlier
studies.

Given the ambiguity of identifications through probability
considerations, the optical faintness of the counterparts, and the
absence of optical lines in particular redshift ranges, it has been
very difficult to estimate the redshift distribution accurately. Where
spectroscopic redshifts cannot be measured for large samples of SMGs,
deep panchromatic surveys can provide photometric redshifts, however
it has not been obvious whether common photometric redshift templates
could be applied indiscriminately to all SMG
counterparts. Furthermore, it has been shown using submm
interferometry with the Atacama Large Millimeter/submillimetre Array
(ALMA)\footnote{www.almaobservatory.org} that the radio/mid-infrared
identification process misses $\sim45$\% of SMGs, and of those it
claims to find, approximately one-third are incorrect
(\citealt{2013ApJ...768...91H}). Additionally, some of the published
redshift distributions derived from different surveys with different
instruments appear to be slightly inconsistent with each other (see
Section 2), which adds more uncertainty to the redshift distribution
of this population of galaxies.

In this paper, we study the impact that selection effects have on
redshift distributions. In Section 2, we summarize the differences
among some of the published redshift distributions of the SMG
population. In Section 3, we describe the simulations that we have
conducted in order to estimate the selection effects (wavelength,
depth, and angular resolution of each survey). In Section 4, we present
our results derived from the simulations and finally in Section 5, we
summarize and discuss our results.

All calculations assume a $\Lambda$ cold dark matter cosmology with
$\Omega_\Lambda=0.68$, $\Omega_m=0.32$, and
$H_0=67.3$kms$^{-1}$Mpc$^{-1}$ (\citealt{2013arXiv1303.5076P}).

\section{Published Redshift distributions}\label{zdists}
The redshift distribution of SMGs (and thus their
cosmic evolution) is not yet completely understood. A comparison of
the redshift distributions from different surveys illustrates this
point.

$\bullet$ \citet{2005ApJ...622..772C} obtained optical spectroscopic
redshifts using the Keck I telescope for a sample of 73 SMGs,
with a median 850 $\mu$m flux density of 5.7 mJy, for which
precise positions were obtained through deep Very Large Array radio
observations. The galaxies lie at redshifts out to $z=3.6$, with a
median redshift of $z_{med}=2.2\pm0.1$. Furthermore, modelling a purely
submm flux-limited sample, based on the expected selection
function for their radio-identified sample, they derived a median
redshift of 2.3. The parent sample of SMGs used for this survey
consists of 150 sources detected at 850 $\mu$m (S/N $>3.0$) with the
Submillimetre Common-User Bolometer Array (SCUBA) on the James Clerk
Maxwell Telescope (JCMT, $\theta\approx14.5$ arcsec) in seven separate
fields with a median depth of $\sigma_{850}\sim1.9$ mJy/beam
(according to the median flux density limit of the subset).

$\bullet$ \citet{2011MNRAS.415.1479W} derived photometric redshifts
from 17 optical to mid-infrared photometric bands for 78 robust
radio, 24 $\mu$m and {\it Spitzer}/Infrared Array Camera (IRAC)
counterparts to 72 of the 126 SMGs selected at 870 $\mu$m (S/N $>3.7$) by the Large
APEX Bolometer Camera (LABOCA) Extended {\it Chandra Deep Field}--South
 Submillimetre Survey (\citealt{2009ApJ...707.1201W})
on the Atacama Pathfinder EXperiment (APEX) 12-m telescope
($\theta\approx19$ arcsec, $\sigma_{870}\sim1.2$ mJy/beam). The median
photometric redshift of the identified SMGs is $z_{med}=2.2\pm0.1$ with
$\sim15$\% high-redshift ($z\ge3$) SMGs. However, a
statistical analysis of sources in the error circles of unidentified
SMGs reveals a population of possible counterparts, which added to the identified SMGs  shifts
the median redshift to $z_{med}= 2.5\pm0.2$.

$\bullet$ \citet{2012MNRAS.420..957Y} reported a redshift distribution
with median redshift of $z_{med}\approx2.6$ with a significant
high-redshift tail of $\sim20$\% at $z\ge3.3$ for 78 SMGs 
detected with AzTEC at 1.1 mm (S/N $>3.5$) in the Great Observatories Origins Deep
Survey-South (GOODS-S) on the Atacama Submillimetre Telescope
Experiment (ASTE), ($\theta\approx30$ arcsec, $\sigma_{1.1}\sim0.6$
mJy/beam; \citealt{2010MNRAS.405.2260S}), and the Great Observatories
Origins Deep Survey-North (GOODS-N) on JCMT ($\theta\approx18$ arcsec,
$\sigma_{1.1}\sim1.0$ mJy/beam, \citealt{2008MNRAS.391.1227P}). The
photometric redshifts were derived by analysing the SEDs
obtained from deep radio continuum, {\it
  Spitzer}/MIPS and IRAC, and LABOCA 870 $\mu$m data, and
complementing the sample with a subset of sources with available
spectroscopic redshifts.

$\bullet$ Using ALMA observations, \citet{2013ApJ...767...88W} found a
median redshift of $z_{med}=3.4$ (taking all ambiguous sources to be
at their lowest redshift option) for a survey of 26 strongly lensed
dusty star-forming galaxies selected with the South Pole Telescope
($\theta\approx1$ arcmin, $\sigma_{1.4}\sim4$ mJy/beam) at 1.4 mm. The
sources were selected to have S$_{1.4mm}>20$ mJy and a dust-like
spectrum and, in order to remove low-$z$ sources, not have bright radio
(S$_{843 MHz} < 6$ mJy) or FIR counterparts (S$_{100 \mu m} <
1$ Jy, S$_{60 \mu m} < 200$ mJy). They found one or more spectral
features in 23 sources yielding an $\sim90$\% detection rate for this
survey with a secure redshift for $\sim70$\% of the sample. Studying
the magnification factors, the sample is expected to cover intrinsic
flux densities of S$_{1.4mm}=1.0-3.0$ mJy.

$\bullet$ \citet{2013arXiv1308.4443R} present the photometric redshift
of 450--$\mu$m--selected sources (S/N $>4.0$), showing a broad peak in the redshift
range $1<z<3$, and a median of $z_{med}=1.4$, combining SCUBA-2
photometry, {\it Herschel}/Spectral and Photometric Imaging Receiver
data from the {\it Herschel} Multi-tiered Extragalactic Survey
, {\it Spitzer}, and {\it Hubble Space Telescope} Wide
Field Camera 3 photometry. The sample consists of sources
detected in the SCUBA-2 Cosmology Legacy Survey
(S2CLS)\footnote{http://www.jach.hawaii.edu/JCMT/surveys/Cosmology.html}
conducted with the JCMT ($\theta\approx7.5$ arcsec) combining observations
on the Ultra Deep Survey (UDS) and the Cosmological Evolution Survey
(COSMOS; \citealt{2013MNRAS.432...53G}) fields to a typical central depth of
$\sigma\approx1.5$ mJy/beam (although the noise is increasing in the radial direction).

$\bullet$ \citet{2013arXiv1302.2619C} derived optical/near-infrared
redshift distributions in the COSMOS field that peak at
$z_{med}=1.95\pm0.19$ for 450 $\mu$m-selected galaxies and
$z_{med}=2.16\pm0.11$ for 850 $\mu$m-selected galaxies (S/N $>3.6$). The two
samples occupy similar areas of the parameter space in redshift and
luminosity, while their median SED peak-wavelengths differ by an
amount consistent with a difference in dust temperature 
$\Delta T_{dust}=8-12$ K. The galaxies were extracted from deep
($\sigma_{450}\approx 4.13$ mJy/beam and $\sigma_{850}\approx 0.80$
mJy/beam ) observations with SCUBA-2/JCMT ($\theta\approx7.5$ and
$\theta\approx14.5$ arcsec for 450 and 850 $\mu$m, respectively).

$\bullet$ Finally, \citet{2013arXiv1311.1485S} presented the first
deep ($\sigma_{2.0}\approx 0.14$ mJy/beam in the central region) map
at 2 mm wavelength centred on the {\it Hubble Deep Field} using the
Goddard-IRAM Superconducting 2 Millimeter Observer (GISMO) at the IRAM
telescope ($\theta\approx17.5$ arcsec). The median redshift of the seven
sources (S/N $>3.0$) with counterparts of known redshifts is $z_{med}=2.91\pm0.94$.

As we can see, the redshift distributions derived from different
surveys achieved with different telescopes show significant
differences. \citet{2012MNRAS.420..957Y} conducted a
Kolmogorov-Smirnov test between their redshift distribution, 
\citet{2005ApJ...622..772C} distribution, and
the 850--$\mu$m--selected \citet{2007MNRAS.379.1571A} distribution, finding that 
the \citeauthor{2005ApJ...622..772C} distribution is substantially
different from the \citeauthor{2012MNRAS.420..957Y} and
\citeauthor{2007MNRAS.379.1571A} distributions. In the same way,
\citet{2009MNRAS.398.1793C} claim that their median redshift of 28 1.1
--mm--selected galaxies of $z_{med}= 2.7$ is statistically distinct from
the $z_{med}= 2.2$ measured in \citeauthor{2005ApJ...622..772C} sample. 
Therefore, it is important to study the impact that selection
effects might have on these measured redshift distributions, in order to
analyse the origin of the differences found.

\section{Simulations}

\begin{figure*}
\includegraphics[width=170mm]{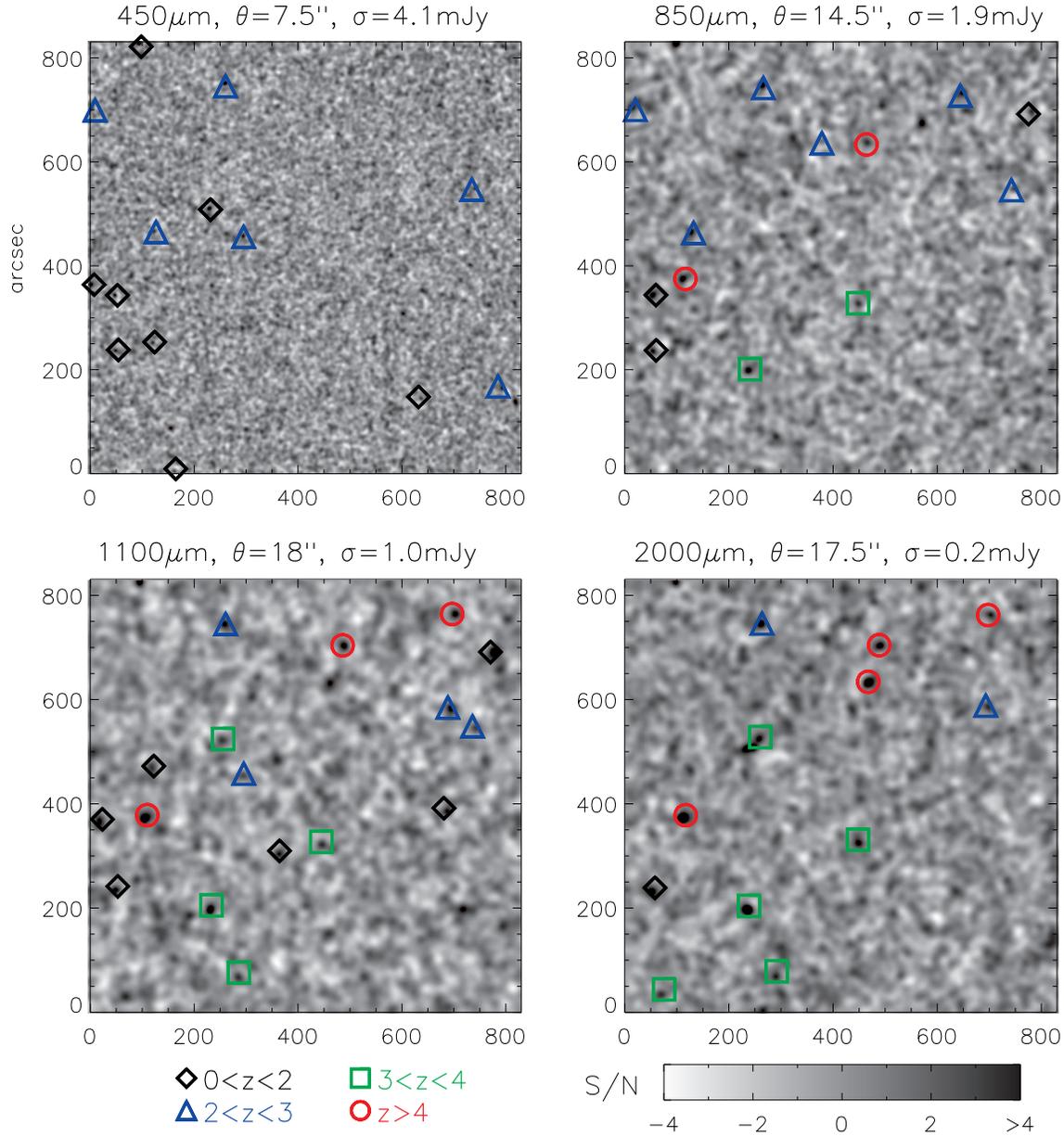}
\caption{An example of the simulated maps (small region of 800 arsec $\times$ 800 arcsec) 
at different wavelengths, depths and angular resolutions, mocking
observations at 450 and 850 $\mu$m with SCUBA-2/JCMT, 1.1 mm with AzTEC/JCMT, 
and 2 mm with GISMO/IRAM. The detected sources (S/N$>3.5$) are represented by
different symbols, where each symbol (and colour, for the online version)  represents 
the redshift range of the detected galaxy --diamond (black) for $0<z\le2$; triangle 
(blue) for $2<z\le3$; square (green) for $3<z\le4$; and circle (red) for $z>4$. The 
characteristics of each map, wavelength, angular resolution (FWHM), and depth, are 
displayed at the top of each panel. }
\label{mapas}
\end{figure*}

In this section we simulate maps with properties similar to those 
from which the published redshift distributions have been derived in order 
to understand the selection effects (i.e. wavelength of observations, angular 
resolution and depth of the maps). We use an  a priori redshift 
distribution and compare the a posteriori redshift distribution 
recovered from each simulated map.

We adopt the source counts at 1.1 mm from \citet{2012MNRAS.423..575S}
derived from 1.6 deg$^2$ blank-field surveys with AzTEC, and the
redshift distribution measured by \citet{2012MNRAS.420..957Y} for 78
SMGs detected with AzTEC at 1.1 mm in the GOODS fields (see Section
\ref{zdists}). We decide to use these priors because 1.1 mm 
 is the central wavelength of surveys considered ($450 \mu$m$ - 2$mm), and 
hence it has a better overlap with the population of galaxies
detected at other wavelengths. Although there are galaxies detected at other
wavelengths that are not detected at 1.1 mm, the conclusion of this paper that 
the 1.1 mm distribution gives a good approximation to the median of a common parent
distribution does not change, for the range considered in this paper ($450 \mu$m$ - 2$mm).

Once we generate maps at 1.1 mm, we can calculate the
maps at different wavelengths using a modified blackbody and the
redshift for each source. The detectable flux density at an observed
frequency $\nu$ from a galaxy with luminosity $L_\nu$ at
redshift $z$ is
\begin{equation}
  S_\nu=\frac{1+z}{4\pi D^2_L}\frac{L_{\nu(1+z)}}{L_\nu},
\end{equation}
where $D^2_L$ is the luminosity distance. We adopt a modified
blackbody, as the SED of each simulated galaxy, with a 
temperature distribution that follows the temperature--luminosity relationship
parametrized by \citet{2012ApJ...761..140C} (see also \citealt{2005ApJ...622..772C} 
and \citealt{2009MNRAS.393..653C}), and alternatively, a random temperature drawn from
a Gaussian distribution with $\langle T\rangle=42 \pm 11$ K, as those measured in deep 
submm surveys (\citealt{2013arXiv1308.4443R}). For the spectral index we use 
a Gaussian distribution with $\langle\beta\rangle=1.6 \pm 0.5$ (\citealt{2013arXiv1308.4443R}).

We co-add this signal map with a noise map where the noise is
represented by a Gaussian with a mean of zero mJy/beam and a standard
deviation equal to the $1\sigma$ depth of each survey described in
Section 2. In the case of the sample of \citet{2012MNRAS.420..957Y},
where the galaxies have been selected from two different surveys, we
have adopted the mean depth of both surveys. On the other hand, in the
case of \citet{2013ApJ...767...88W}, where the sample consists of
lensed galaxies, we have adopted a noise ($1\sigma=0.6$ mJy) such that the
galaxies detected (S/N$>3.5$) in our simulated map have a flux density
similar to the mean delensed (intrinsic) flux density estimated by
\citet{2013ApJ...767...88W}. Finally, in the case of \citet{2013arXiv1308.4443R},
we have simulated a noise map in which the noise increases radially from 
$\sigma\approx1.5$ mJy/beam to $\sigma\approx5$ mJy/beam on the edges, similar 
to the  noise properties in the daisy maps of S2CLS. The remaining surveys
are quite uniform and  do not need a special treatment of their noise properties.

Finally, we convolve this co-added map with a Gaussian point spread
function (PSF) with a full width at half--maximum (FWHM) equal to the
angular resolution of the different surveys we want to simulate. In
Fig. \ref{mapas}, we show an example of a small field within our
simulated maps at different wavelengths, different depths, and
different angular resolutions generated with this procedure, and the
galaxies detected at S/N$>3.5$ in each map. In this example, we show
maps at 450 $\mu$m, 850 $\mu$m, 1.1 mm, and 2 mm with angular
resolution and depth similar to the surveys published by
\citet{2013arXiv1302.2619C}, \citet{2005ApJ...622..772C},
\citet{2008MNRAS.391.1227P}, and \citet{2013arXiv1311.1485S},
respectively (see Section 2).

In each realization, we simulate seven maps of 400 sq. arcmin that
reproduce the wavelength, depth, and angular resolution of the surveys
that we wish to analyse. The size of the maps allows us enough 
statistics in each realization. Once we have all the simulated
maps, the next step is to extract the sources and to calculate the
redshift distribution of each map. A source is considered to be
recovered in each map if it is detected with S/N $>\xi_{tresh}$ within a
search radius equal to the FWHM of the beam, where $\xi_{tresh}$ is the 
value used in each of the surveys described in Section 2 ($\xi_{tresh}=3-4$).
We conduct 100 realizations to estimate the redshift distribution extracted for each
map, the median value, and the error on the median calculated as the
standard deviation of the median values in each realization.

\section{Results}

\subsection{The redshift distribution}

\begin{figure}
\includegraphics[width=90mm]{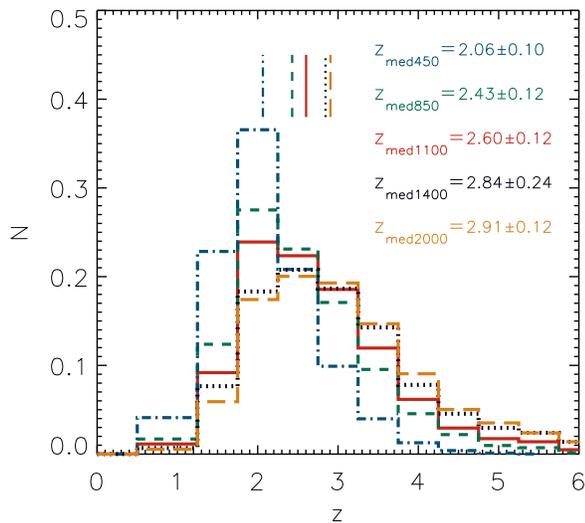}
\caption{Extracted redshift distributions from the simulated 450 
$\mu$m--shallow map (dash--dotted blue line),  850 $\mu$m--map 
(short-dashed green line),  1.1 mm--map (solid red line), 1.4 mm--map 
(dotted black line), and 2 mm--map (long-dashed orange line). Top bars 
represent the corresponding median values of the distributions.}
\label{histogramas}
\end{figure}

Although we have only one a priori redshift distribution, the
extracted redshift distributions from the simulated maps have
different shapes (see Fig. \ref{histogramas}). The median redshift varies
increasingly from $z_{med}=2.06\pm0.10$ to $2.91\pm0.12$ as
the wavelength of the simulated maps changes from 450 $\mu$m to 2 mm
respectively. This result does not depend on whether we assume that the temperature
of SMGs follows the same Gaussian distribution for all sources or the temperature-luminosity 
relationship. We will call this set of redshift distributions,
hereafter, the {\it expected} redshift distributions, since 
these are the distributions we expect to measure after taking into
account selection effects. We will compare these expected redshift
distributions with the published distributions.

\citet{2013arXiv1302.2619C} derived a redshift distribution with a
median of $z_{med}=1.95 \pm 0.19$ for their 450 $\mu$m--map
($\sigma\sim4.1$ mJy); in our simulations, we estimate for this survey a
redshift distribution with $z_{med}=2.06 \pm 0.10$, consistent with
the published value. This means that we are missing some high-redshift
galaxies from the original parent distribution ($z_{med}\approx2.6$;
\citealt{2012MNRAS.420..957Y}) through selection effects and for this
reason the median is shifted towards lower redshifts.

\citet{2005ApJ...622..772C} estimated a redshift distribution with
$z_{med}=2.2 \pm 0.2$, whilst we have derived from the simulated map a
distribution with $z_{med}=2.43 \pm 0.12$ which is compatible within
the error bars with the published value. \citeauthor{2005ApJ...622..772C} 
proposed a corrected model for their distribution that takes into account 
the expected radio bias, and suggested a revised $z_{med}\sim2.3$, in 
better agreement with our extracted distribution. Hence the redshift
distribution of \citeauthor{2005ApJ...622..772C} is also consistent with 
our a priori redshift distribution.

\citet{2011MNRAS.415.1479W} derived $z_{med}=2.2 \pm 0.1$ from their
870 $\mu$m--survey. A statistical analysis of the unidentified sources,
however, shifts the distribution to $z_{med}\approx2.5 \pm 0.3$ in
very good agreement with our expected $z_{med}=2.46 \pm
0.10$. Moreover, a recent study of the same submillimetre source sample
with ALMA, which ensures the photometric redshift of the correct counterparts 
are used (\citealt{2013arXiv1310.6363S}) finds the same result with a 
$z_{med}=2.5\pm0.2$ after correcting for incompleteness.

\citet{2013arXiv1311.1485S} reported a median redshift of $z_{med}=2.91\pm0.94$
at a wavelength of 2 mm, in very good agreement with the 
value that we have extracted from our simulated map ($z_{med}=2.91\pm0.12$).

The \citet{2012MNRAS.420..957Y} distribution is obviously in agreement
with our expected distribution, since it is the one we have adopted
as the a priori distribution. This confirms that our simulations
and the source extraction method are working correctly, since we are 
recovering the same distribution at this wavelength.

The photometric redshift distribution
of \citet{2012A&A...548A...4S}, who found a median redshift of
$z_{med}=3.1\pm0.3$ for 17 galaxies detected at 1.1 mm in the COSMOS
field, is also within the 68\% confidence level bars of
our value, but there is an offset from the \citet{2012MNRAS.420..957Y}
distribution. The offset could be due in part to cosmic variance, 
where the COSMOS field is known to have several notable, very distant $z>4.5$
galaxies (e.g. \citealt{2008ApJ...681L..53C}; \citealt{2010ApJ...720L.131R}) as
discussed by \citet{2014arXiv1402.1456C}.

However, there are two distributions which at first are not compatible with our
expected redshift distribution for these surveys; these are as follows:

a) The \citet{2013arXiv1308.4443R} redshift distribution, derived from
their deep 450 $\mu$m--map, has a median of $z_{med}=1.4$, which is in
disagreement with our expected distribution ($z_{med}=2.13\pm0.08$). 
It is important to remark the differences between the \citet{2013arXiv1308.4443R}
and the \citet{2013arXiv1302.2619C} ($z_{med}=1.95\pm0.19$) redshift
distributions because they were extracted from surveys at the same
wavelength and angular resolution, but, they peak at different
redshifts. This offset is likely due to the difference in depths
between both surveys as discussed by \citet{2014arXiv1402.1456C}. 
The work by \citeauthor{2013arXiv1308.4443R} is
a factor of $\sim 3$ deeper and therefore they were able to detect
fainter galaxies at lower redshift. In conclusion, in order to explain the
\citeauthor{2013arXiv1308.4443R} distribution we would need to consider
another population of galaxies which lie at lower redshifts and
probably with different luminosities and SFRs (i.e. Luminous Infrared Galaxies, LIRGs) than
longer--wavelength selected SMGs. The properties that this population of 
galaxies should have in order to reproduce the redshift distribution 
are beyond the scope of this paper.

b) The \citet{2013ApJ...767...88W} redshift distribution is also in
disagreement with our expected distribution, since our median redshift
is $z_{med}=2.84 \pm 0.24$. However, as \citeauthor{2013ApJ...767...88W} noted, 
their selection of bright 1.4 mm sources imposes a requirement that they be
gravitationally lensed, effectively suppressing sources at $z<1.5$ due
to the low probability of being lensed at these redshifts. Since our
a priori redshift distribution has a significant proportion of
sources with $z<1.5$ ($\sim5 \%$, and $\sim20 \%$ with $z<2$, where the
effect is still significant) we do not expect to have compatible
distributions. In order to analyse the lensing bias, we have removed the 
low-redshift sources from our extracted redshift distribution according to the 
probability of strong lensing as a function of redshift. 
In order to estimate this probability, we used a mean source magnification of $\mu=10$ 
(the sample has magnifications of $\mu= 5-21$) and normalized the lensing probability 
estimated by \citet{2013ApJ...767...88W} for this magnification, at $z=4$, where the 
probability becomes flat.  Once we have removed the low-redshift sources due to the 
strong lensing effect, the extracted redshift distribution has a median redshift of  
$z_{med}=3.09 \pm 0.21$, which is in agreement within the error bars with the derived 
$z_{med}=3.4 \pm 0.25$. We have  estimated the error in the \citeauthor{2013ApJ...767...88W}
median redshift  using a bootstrapping method, and taking all ambiguous sources to be
at their lowest redshift solution. \citet{2013arXiv1310.6363S} also reached the same 
conclusion when comparing their 850--$\mu$m--selected redshift distribution and the 
lensing--corrected 1.4--mm--selected redshift distribution.

Aditionally, the galaxies observed by \citeauthor{2013ApJ...767...88W} could be 
intrinsically brighter than the average SMGs and therefore could lie at higher 
redshifts as suggested by \citet{2013arXiv1312.1173K}.

We have thus shown that the measured distributions are mostly
consistent with a parent distribution with $z_{med}\approx2.6$, very
similar to the redshift distribution of \citet{2012MNRAS.420..957Y} when taking
into account wavelength selection, depth, and angular resolution
limitations. We should stress that  we are not considering 
the difficulties intrinsic to finding multiwavelength counterparts for these 
sources, or obtaining spectroscopic redshift which could bias the measured
redshift distributions observationally. These biases are hence intrinsic to
this study and to the adopted prior.

On the other hand, we are also interested in determining which of
these parameters is the most crucial factor in imposing
differences on the measured redshift distributions. In order to
quantify this, we have followed the same procedure as described in 
Section 3, but now, with an angular resolution of $\theta=1$ arcsec
(comparable to ALMA resolutions) for all the simulated maps. 

\begin{figure}
\includegraphics[width=90mm]{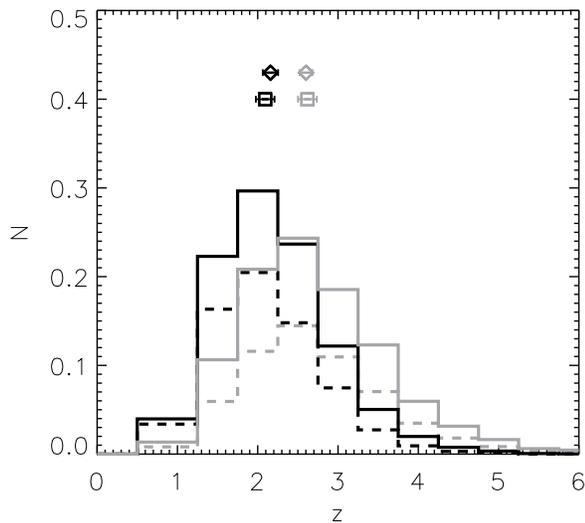}
\caption{Extracted redshift distributions from the simulated 450 $\mu$m-shallow 
map  with a resolution of $\theta=1''$ (solid black line) and $\theta=7.5''$ 
(dashed black line), and from the simulated 1.1 mm-map with a resolution of 
$\theta=1''$ (solid gray line) and $\theta=19''$ (dashed gray line).  The median 
and the error on the median for each distribution (square and diamond for the 
poorer and better angular resolution maps respectively) are plotted at the top of 
the graph. The histograms have been scaled by the total number of galaxies extracted 
from the map with the highest angular resolution, in each wavelength.}
\label{almaresolution}
\end{figure}

With this high angular resolution the blending of sources is
insignificant and therefore we are able to recover a large fraction of
sources in each map. However, the
redshift distribution extracted from each simulated map is very similar 
to the one we measured from the poorer angular
resolution maps. In Fig. \ref{almaresolution} we show the
extracted redshift distribution at 450 $\mu$m and 1.1 mm for both,
poor ($\theta=7.5$ and $\theta=19$ arcsec respectively) and best
($\theta=1$ arcsec) angular resolutions. At each wavelength, the histograms
have been scaled by the total number of galaxies extracted from the
map with the highest angular resolution, in order to compare them
easily.

The similarity of the histograms confirms that wavelength and depth mainly determine the 
redshift distribution, and therefore the angular resolution is only a secondary
effect in imposing a bias. This result is in agreement with the work 
of \citet{2013arXiv1310.6363S}, who did an ALMA follow-up of the
\citet{2011MNRAS.415.1479W} sample finding largely the same redshift distribution with 
the new high-resolution data. However, the fainter population of  galaxies that has 
not yet been detected with single-dish telescopes (due to the confusion limit) could have 
a different redshift distribution, and therefore the angular resolution could be an important
effect at these smaller flux densities.

\subsection{The SED temperature}

We have shown that shorter wavelength--maps miss high redshift galaxies. 
It is also important to understand the
differences between the subsets of galaxies selected at different
wavelengths. As we have the temperature and
emissivity index for each recovered galaxy (grey--body distribution,
see Section 3) we can investigate the properties of the SEDs of the
galaxies extracted in the different simulated maps. Analysing these
properties, we have found that the main difference between the
galaxies selected at different wavelengths is the grey--body
temperature, where the shorter-wavelength maps select a hotter
population. The mean SED peak wavelength monotonically shifts from
$109\pm 4 \mu$m to $123 \pm 3$ (or $T=46\pm2$ to $T=40\pm1$ K) 
for 450 $\mu$m--selected galaxies to 2 mm--selected galaxies, 
respectively. There is no significant difference between the 1.1 mm, 
870 $\mu$m, and 850 $\mu$m maps, which have a mean  SED peak of $118\pm 4 \mu$m 
(or $T\approx42\pm2$ K).

This effect has been previously discussed  by \citet{2013arXiv1302.2619C}, 
where they also found that their two samples (SCUBA-2 450 and 850 $\mu$m-selected 
galaxies in the COSMOS field) occupy a similar parameter space in redshift and
luminosity, with a difference in the SED peak of $20-50\mu$m or a temperature 
difference of $\Delta T_{dust}=8-12$ K, consistent, within the 
error bars, with our simulations.

\subsection{The K--S test}

A common method to compare distributions is the Kolmogorov--Smirnov
(K-S) test which tells us the probability that two distributions are
drawn from the same parent distribution. We have applied this test to
all the pairs of redshift distributions extracted from the different
simulated maps.
In order to compare with the results from previous
studies we have chosen 70 random sources for each simulated map and then we have applied the K-S test
to these subsamples. The results are summarized in Table 1.

As can be seen in Table 1, we can reject with better than 99 per
cent confidence the hypothesis that the subsample of 450--$\mu$m--selected 
galaxies has a common parent redshift distribution with 
the subsamples extracted from longer--wavelength simulated maps, even when we know
that all these distributions have been generated adopting the same parent distribution. The
same hypothesis for other wavelengths cannot be rejected with such a high confidence.  
Hence, if we
want to know that if two distributions extracted from surveys carried out at different wavelengths 
are compatible with a common
parent distribution, one should implement simulations similar
to those described here, since the K--S test is not able to
take into account the selection effects introduced by the choice of wavelength and depth.

\begin{table}
\caption{K-S test probability that two redshift distributions are drawn from the same parent 
distribution. A subset of 70 galaxies has been randomly selected from each simulated map.}
\begin{tabular}{cccccc}
\hline
&450  $\mu$m&  850 $\mu$m& 1.1 mm&1.4 mm&2.0 mm\\
\hline
450  $\mu$m& -- & 0.001&3.61e--4 &4.43e--7&1.73e--4\\
850 $\mu$m&  & --& 0.857& 0.069&0.106\\
1.1 mm &  & & --&0.326&0.443\\
1.4 mm & & & & -- &0.443\\
\hline
\end{tabular}
\label{tabla1}
\end{table}

\section{Discussion and Conclusions}

We have analysed the selection effects that wavelength, depth, and angular
resolution impose on the extracted redshift distributions from different
surveys. We have found that some of the published redshift
distributions, which were reported to
be inconsistent with each other, are in agreement with a common parent
distribution. The differences between these published redshift distributions 
can be explained by selection effects imprinted mainly by differences in 
wavelength and depth of the observations. 

The median redshifts derived from our simulations ($z_{med}=2.06-2.91$)
are in very good agreement with the values previously reported
(\citealt{2005ApJ...622..772C}; \citealt{2011MNRAS.415.1479W};
\citealt{2012MNRAS.420..957Y}; \citealt{2013arXiv1302.2619C};
\citealt{2013arXiv1311.1485S}), which indicate the consistency between
the published redshift distributions and our adopted parent
distribution, even when some of these distributions have been shown to be
statistically inconsistent with each other (\citealt{2009MNRAS.398.1793C};
\citealt{2012MNRAS.420..957Y}). We can also reproduce with the same 
parent distribution the redshift distribution of \citet{2013ApJ...767...88W} 
which peaks at $z_{med}=3.4$, when we take into account the bias imposed by the 
lensing probability. We conclude that in order to test  compatibility of this
kind of distributions, which have been extracted from surveys with different selection
effects, the best way is to use  simulations (similar to the procedure described here).

As expected, and previously described by \citet{2013arXiv1302.2619C}, the main 
difference between the galaxies selected at different wavelengths is the SED 
temperature. The mean SED peak wavelength shifts from $109 \pm4 \mu$m to 
$123 \pm3 \mu$m (or T$=46 \pm2$ to T$=40\pm1$) from 450 $\mu$m--selected galaxies to
2 mm--selected galaxies.

Finally, as we have shown here, a comprehensive view and accurate
determination of the redshift distribution of SMGs need to be based on the complementarity of
multi--wavelength observations from large statistically significant
samples. Future multiwavelength large-format cameras, like those designed to operate
at the Large Millimeter Telescope, CCAT, and ALMA, will contribute towards this goal as they target similar 
cosmological fields.

\section*{Acknowledgments}
We would like to thank an anonymous referee for a constructive and helpful report,
which has significantly improved this paper.
This work has been mainly supported by Mexican CONACyT research grants
CB-2011-01-167291 and CB-2009-133260. This work was also supported in part by the 
National Science Foundation under Grant No. PHYS-1066293 and the hospitality of the 
Aspen Center for Physics. IA would like to thank the participants of the workshop
``The Obscured Universe: Dust and Gas in Distant Starburst Galaxies'' for discussions and 
suggestions on this topic.

\bsp

\label{lastpage}


\begin{thebibliography}{}
%
\bibitem[\protect\citeauthoryear{Aretxaga et al.}{2007}]{2007MNRAS.379.1571A} Aretxaga, I., Hughes,  D. H., Coppin, K., et al., 2007, MNRAS, 379, 1571 

\bibitem[\protect\citeauthoryear{Blain \& Longair}{1993}]{1993MNRAS.264..509B} Blain, A.~W., \& Longair, M.~S., 1993, MNRAS, 264, 509 

\bibitem[\protect\citeauthoryear{Blain et al.}{2002}]{2002PhR...369..111B} Blain, A.~W., Smail, I., Ivison, R.~J., Kneib, J.-P., \& Frayer, D.~T., 2002, Phys. Rep., 369, 111 

\bibitem[\protect\citeauthoryear{Capak et al.}{2008}]{2008ApJ...681L..53C} Capak, P., Carilli, C.~L., Lee, N., et al., 2008, ApJ, 681, L53 

\bibitem[\protect\citeauthoryear{Casey et al.}{2012}]{2012ApJ...761..140C} Casey, C.~M., Berta, S.,  B{\'e}thermin, M., et al., 2012, ApJ, 761, 140 

\bibitem[\protect\citeauthoryear{Casey et al.}{2013}]{2013arXiv1302.2619C} Casey, C.~M., Chen, C.-C., Cowie, L.~L., et al., 2013, MNRAS, 436, 1919 

\bibitem[\protect\citeauthoryear{Casey et al.}{2014}]{2014arXiv1402.1456C} Casey, C.~M., Narayanan, D., \& Cooray, A., 2014, preprint (arXiv:1402.1456) 

\bibitem[\protect\citeauthoryear{Chapin et al.}{2009}]{2009MNRAS.398.1793C} Chapin, E.~L., Pope, A., Scott, D., et al., 2009, MNRAS, 398, 1793 

\bibitem[\protect\citeauthoryear{Chapin et al.}{2009}]{2009MNRAS.393..653C} Chapin, E.~L., Hughes, D.~H., \& Aretxaga, I., 2009, MNRAS, 393, 653 

\bibitem[\protect\citeauthoryear{Chapman et al.}{2005}]{2005ApJ...622..772C} Chapman, S.~C., Blain, A.~W., Smail, I., \& Ivison, R.~J., 2005, ApJ, 622, 772 

\bibitem[\protect\citeauthoryear{Chen et al.}{2013}]{2013ApJ...762...81C} Chen, C.-C., Cowie, L.~L., Barger, A.~J., et al., 2013, ApJ, 762, 81 

\bibitem[\protect\citeauthoryear{Dunne et al.}{2000}]{2000MNRAS.319..813D} Dunne, L., Clements, D.~L., \& Eales, S.~A., 2000, MNRAS, 319, 813 

\bibitem[\protect\citeauthoryear{Dunne \& Eales}{2001}]{2001MNRAS.327..697D} Dunne, L., \& Eales, S.~A., 2001, MNRAS, 327, 697 

\bibitem[\protect\citeauthoryear{Geach et al.}{2013}]{2013MNRAS.432...53G} Geach, J.~E., Chapin, E.~L., Coppin, K.~E.~K., et al., 2013, MNRAS, 432, 53 

\bibitem[\protect\citeauthoryear{Hodge et al.}{2013}]{2013ApJ...768...91H} Hodge, J.~A., Karim, A., Smail, I., et al., 2013, ApJ, 768, 91 

\bibitem[\protect\citeauthoryear{Hughes et al.}{1998}]{1998Natur.394..241H} Hughes, D.~H., Serjeant, S., Dunlop, J., et al., 1998, Nature, 394, 241 

\bibitem[\protect\citeauthoryear{Iono et al.}{2006}]{2006ApJ...640L...1I} Iono, D., Peck, A.~B., Pope, A., et al., 2006, ApJ, 640, L1 

\bibitem[\protect\citeauthoryear{Koprowski et al.}{2013}]{2013arXiv1312.1173K} Koprowski, M.~P., Dunlop, J.~S., Michalowski, M.~J., et al., 2013, arXiv:1312.1173 

\bibitem[\protect\citeauthoryear{Oliver et al.}{2010}]{2010A&A...518L..21O} Oliver, S.~J., Wang, L., Smith, A.~J., et al., 2010, A\&A, 518, L21 

\bibitem[\protect\citeauthoryear{Perera et al.}{2008}]{2008MNRAS.391.1227P} Perera, T.~A., Chapin, E.~L., Austermann, J.~E., et al., 2008, MNRAS, 391, 1227 

\bibitem[\protect\citeauthoryear{Planck Collaboration et al.}{2013}]{2013arXiv1303.5076P} Planck Collaboration, Ade, P.~A.~R., Aghanim, N., et al., 2013, arXiv:1303.5076 

\bibitem[\protect\citeauthoryear{Riechers et al.}{2010}]{2010ApJ...720L.131R} Riechers, D.~A., Capak, P.~L., Carilli, C.~L., et al., 2010, ApJ, 720, L131 

\bibitem[\protect\citeauthoryear{Roseboom et al.}{2013}]{2013arXiv1308.4443R} Roseboom, I.~G., Dunlop, J.~S., Cirasuolo, M., et al., 2013, MNRAS, 436, 430 

\bibitem[\protect\citeauthoryear{Scott et al.}{2010}]{2010MNRAS.405.2260S} Scott, K.~S., Yun, M.~S., Wilson, G.~W., et al., 2010, MNRAS, 405, 2260 

\bibitem[\protect\citeauthoryear{Scott et al.}{2012}]{2012MNRAS.423..575S} Scott, K.~S., Wilson, G.~W., Aretxaga, I., et al., 2012, MNRAS, 423, 575 

\bibitem[\protect\citeauthoryear{Simpson et al.}{2014}]{2013arXiv1310.6363S} Simpson, J.~M., Swinbank, A.~M., Smail, I., et al., 2014, ApJ, 788, 125

\bibitem[\protect\citeauthoryear{Smail et al.}{1997}]{1997ApJ...490L...5S} Smail, I., Ivison, R.~J., \& Blain, A.~W., 1997, ApJ, 490, L5 

\bibitem[\protect\citeauthoryear{Smol{\v c}i{\'c} et al}{2012}]{2012A&A...548A...4S} Smol{\v c}i{\'c}, V., Aravena, M., Navarrete, F., et al., 2012, A\&A, 548, A4 

\bibitem[\protect\citeauthoryear{Staguhn et al.}{2013}]{2013arXiv1311.1485S} Staguhn, J.~G., Kovacs, A., Arendt, R.~G., et al., 2013, preprint (arXiv:1311.1485) 

\bibitem[\protect\citeauthoryear{Wang et al.}{2007}]{2007ApJ...670L..89W} Wang, W.-H., Cowie, L.~L., van Saders, J., Barger, A.~J., \& Williams, J.~P., 2007, ApJ, 670, L89 

\bibitem[\protect\citeauthoryear{Wardlow et al.}{2011}]{2011MNRAS.415.1479W} Wardlow, J.~L., Smail, I., Coppin, K.~E.~K., et al., 2011, MNRAS, 415, 1479 

\bibitem[\protect\citeauthoryear{Wei{\ss} et al.}{2009}]{2009ApJ...707.1201W} Wei{\ss}, A., Kov{\'a}cs, A., Coppin, K., et al., 2009, ApJ, 707, 1201 

\bibitem[\protect\citeauthoryear{Wei{\ss} et al.}{2013}]{2013ApJ...767...88W} Wei{\ss}, A., De Breuck, C., Marrone, D.~P., et al., 2013, ApJ, 767, 88 

\bibitem[\protect\citeauthoryear{Younger et al.}{2007}]{2007ApJ...671.1531Y} Younger, J.~D., Fazio, G.~G., Huang, J.-S., et al., 2007, ApJ, 671, 1531 

\bibitem[\protect\citeauthoryear{Younger et al.}{2009}]{2009ApJ...704..803Y} Younger, J.~D., Fazio, G.~G., Huang, J.-S., et al., 2009, ApJ, 704, 803 

\bibitem[\protect\citeauthoryear{Yun et al.}{2012}]{2012MNRAS.420..957Y} Yun, M.~S., Scott, K.~S., Guo, Y., et al., 2012, MNRAS, 420, 957 





\end{thebibliography}
\end{document}